\documentclass[epj]{svjour}

\usepackage{graphicx}
\usepackage{fancyhdr}
\def\makeheadbox{{
 }}
\newcommand{\alt}{\mathrel{\raisebox{-.6ex}{$\stackrel{\textstyle<}{\sim}$}}}
\newcommand{\agt}{\mathrel{\raisebox{-.6ex}{$\stackrel{\textstyle>}{\sim}$}}}

\def\mytitle{My title} 
\def\myauthors{My name}  
\def\mytype{My type of session}
\def\mysession{My session}

\def\mytitle{
Roles of Higgs decay into two pseudoscalar bosons \\
in the search of intermediate-mass Higgs Boson
} 
\def\myauthors{
Kingman Cheung, Jeonghyeon Song, Qi-Shu Yan}   
\def\mytype{Contributed Talk}    
\def\mysession{Colliders - Higgs Phenomenology}

\begin{document}
\title{
Roles of Higgs decay into two pseudoscalar bosons \\
in the search of intermediate-mass Higgs Boson
}
\author{Kingman Cheung \inst{1,2}
 \and
 Jeonghyeon Song \inst{3}
  \and
 Qi-Shu Yan \inst{1,2}
}                     
%
%
\institute{Department of Physics, National Tsing Hua University,
Hsinchu, Taiwan 
\and 
Physics Division, National Center for Theoretical Sciences, Hsinchu, Taiwan
 \and
Department of Physics, Konkuk University, Seoul 143-701, Korea
}
%
\date{}
\abstract{
The dominance of $h\to \eta \eta$ decay mode for the intermediate mass 
Higgs boson is highly motivated to solve the little hierarchy 
problem and to ease the tension with the precision data. 
However, the discovery modes for $m_h \alt 150$ GeV, 
$h \to \gamma\gamma$ and $W/Z h \to (\ell\nu/\ell \bar \ell) (b\bar b)$, 
will be substantially affected.  
We show that $h \to \eta \eta \to 4b$ is complementary
and we can use this decay mode to detect the intermediate 
Higgs boson at the LHC, via $Wh$ and $Zh$ production.  
Requiring at least one charged lepton and 4 $B$-tags in the final state, we
can identify a clean Higgs boson signal for $m_h \alt 150$ GeV
with a high significance and with a full Higgs mass 
reconstruction.
We use the next-to-minimal supersymmetric standard model and the 
simplest little Higgs model for illustration.
} 
\maketitle
\section{Introduction}
\label{intro}

The standard model (SM) has been successful in explaining all the data,
except for a few observations.  One of them is the controversy between
the precision data and the direct search for the SM Higgs boson.  The
precision measurements from LEP and SLD collaborations strongly prefer
a light Higgs boson with a mass around 100 GeV\,\cite{prec}.
However, the direct search has put a lower bound of 114.4 GeV\,\cite{LEP}.
Such a high Higgs mass bound also induces the so-called little
hierarchy problem in supersymmetric framework.
It is urgent to relieve the tension arised from the Higgs mass bound.  

A phenomenological approach to lower the Higgs mass bound
is to reduce either the coupling $g_{ZZh}$ or $B(h \to b\bar b)$.
One possibility is to add a singlet field to the Higgs
sector such that the Higgs doublet and the singlet mix.
The SM-like Higgs boson will have a smaller effective
coupling $g_{ZZh}$ to the $Z$ boson.
More important is that there are additional decay modes
for the Higgs boson.
In supersymmetric framework, the most popular approach is the
next-to-minimal supersymmetric standard model (NMSSM).
It has been
shown \cite{nmssm} that, in most parameter space that is natural, 
the SM-like Higgs boson can decay into
a pair of light pseudoscalar bosons with a branching ratio larger than $0.7$.
The Higgs mass bound can be as low as around 100 GeV.
In little Higgs framework, it has been shown \cite{song} that
in the simplest little Higgs model with the $\mu$ parameter 
(SLH$\mu$)\,\cite{simple},
the Higgs boson can dominantly decay into a pair of pseudoscalar
bosons $\eta$.  Together with the
reduction of the $g_{ZZh}$ coupling, the Higgs mass bound can be lowered.
In these models, the Higgs boson dominantly decays into lighter
Higgs bosons (we shall denote the lighter Higgs boson as pseudoscalar
boson $\eta$ without loss of generality.)  The dominance of
$h\to\eta\eta$ mode for the intermediate Higgs boson has significant
impacts on the Higgs search strategies.
The most useful channel for intermediate Higgs boson, $h\to \gamma\gamma$,
will be substantially affected because $B(h\to\gamma\gamma)$ lowers by 
a factor of a few.  So is the $h\to b \bar b$ in $Wh,Zh$ production.
It is therefore utmost important to show the complementarity of the 
$h\to \eta\eta$ mode, and timely to establish the feasibility of the 
$h\to \eta \eta$ mode.  
We have shown in Ref. \cite{ours} that
using $h \to \eta \eta \to 4b$ for $m_\eta > 2 m_b$ the Higgs signal 
can be identified at the LHC, via $Wh, Zh$ production.  With
at least one charged lepton and $4 B$-tags in the final state, one
can obtain a clean signal of high significance and a full Higgs mass
reconstruction.

\section{Production and decay}
\label{sec1}

The pseudoscalar boson decays into the heaviest fermion pair that
is kinematically allowed, either $b\bar b$ or $\tau^+ \tau^-$.
If $m_\eta \! > \!2 m_b$,
the SM-like Higgs boson will decay like
$h \to \eta \eta \to (4b, 2b2\tau, 4\tau)$.
Feasibility studies
focusing on Higgs production at the Tevatron
have been performed in extended supersymmetric models.
The $gg \to h \to \eta\eta \to 4b$ signal at the Tevatron has been
shown overwhelmed by large QCD background \cite{scott}.
Similar conclusions can be drawn for the LHC.
Another study using $(2b,2\tau)$ mode for the associated Higgs production
with a $W/Z$ at the Tevatron was performed,
but a full Higgs mass reconstruction is difficult \cite{carena}.
The $4\tau$ mode was also studied at the Tevatron for
$ 2 m_\tau < m_\eta  < 2 m_b$ \cite{4tau}.
If $m_\eta \!<\! 2 m_\mu$, on the other hand, the modes $\eta \to e^+ e^-,
\gamma\gamma$ become dominant \cite{matchev}
but the photon pair for each pseudoscalar decay is
very collimated, which reduces the detectability.
One can also have the pseudoscalar boson produced directly, \textit{e.g.},
in the associated production with a gaugino pair\,\cite{assoc}.

\begin{table*}[tb!]
\centering
\caption{\small \label{table1}
Signal cross sections for $Wh$ and  $Zh$ production for
bench-mark points NMSSM (A) and NMSSM (B), and for
SLH$\mu$ (A) and SLH$\mu$ (B) at the LHC.
}
\smallskip
\begin{tabular}{lllll}
\hline
Channels & NMSSM (A)  & NMSSM (B) & SLH$\mu$ (A) & SLH$\mu$ (B)   \\
\hline
       & $\lambda=0.18,\;\kappa=-0.43$  & $\lambda=0.26,\;\kappa=0.51$
       & $f=4$ TeV & $f=2$ TeV  \\
       & $\tan\beta=29$
       & $\tan\beta=23$
       & $\mu=20$ GeV  & $\mu=20$ GeV \\
       & $A_\lambda=-437$ GeV
       & $A_\lambda=-222$ GeV
       & $x_\lambda=5.86$
       & $x_\lambda=10$ \\
       & $A_\kappa=-4$ GeV & $A_\kappa=-13$ GeV
       & $\tan\beta=17$ & $\tan\beta=9.47$  \\
       & $\mu_{\rm eff}=-143$ GeV & $\mu_{\rm eff}=144$ GeV & & \\
 \hline
       & $m_{h_1} =110$ GeV & $m_{h_1} = 109$ GeV
       & $m_{h} =146.2$ GeV & $m_{h} = 135.2$ GeV \\
       & $m_{a_1} =30$ GeV & $m_{a_1} = 39$ GeV
       & $m_{\eta}=68.6$ GeV & $m_{\eta} = 47.9$ GeV \\
       & $B(h_1\to a_1 a_1) =0.92$ & $B(h_1\to a_1 a_1) = 0.99$
       & $B(h\to \eta\eta) =0.65$ & $B(h\to \eta\eta) =0.75$ \\
       & $B(a_1\to b \bar b) =0.93$ & $B(a_1\to b \bar b) = 0.92$
       & $B(\eta\to b\bar b) =0.85$ & $B(\eta\to b\bar b) =0.86$ \\
   & $g_{VVh_1}/g_{VVh}^{\rm SM} =0.99$ & $g_{VVh_1}/g_{VVh}^{\rm SM} =-0.99$
   & $g_{VVh}/g_{VVh}^{\rm SM} =0.57$ & $g_{VVh}/g_{VVh}^{\rm SM} = 0.44$ \\
   & $g_{tth_1}/g_{tth}^{\rm SM} =0.99$ & $g_{tth_1}/g_{tth}^{\rm SM} =-0.99$
   & $g_{tth}/g_{tth}^{\rm SM} =0.79$ & $g_{tth}/g_{tth}^{\rm SM} =0.93$ \\
       & $g_{tta_{1}}/g_{tth}^{\rm SM} =-2.4 \times 10^{-3}$
       & $g_{tta_{1}}/g_{tth}^{\rm SM} =-1.2 \times 10^{-2}$
    & $g_{tt\eta}/g_{tth}^{\rm SM} =-0.89$
    & $g_{tt\eta}/g_{tth}^{\rm SM} = -1.38$ \\
    & $C^2_{4b} =0.80$ & $ C^2_{4b}=0.83$
    & $C^2_{4b} =0.16$ & $ C^2_{4b}=0.11$ \\
\hline
$W^+h$ signal  & 3.13 fb & 9.54  fb &1.27 fb   & 0.63 fb \\
$W^-h$ signal  & 2.35 fb  & 6.55  fb &0.87  fb  & 0.44 fb \\
$Zh$ signal  & 1.05  fb  & 2.76 fb  &0.36  fb  & 0.18 fb \\
\hline
\end{tabular}
\end{table*}

In Ref. \cite{ours}, we focus on $Wh$ and $Zh$ production at the LHC,
followed by the leptonic decay of the $W$ and $Z$,
and $h \to \eta \eta \to b\bar b b\bar b$.
In the final state, we require a charged lepton
and 4 $b$-tagged jets.  The advantage of having a charged lepton in
the final state is to suppress the QCD background.
We require  $4$ $b$-tagged jets
to avoid the huge $t\bar t$ background.  We are still
left with some irreducible
backgrounds from $W+ n b$ and $Z +  n b$ production with $n \ge 4$, 
$t\bar t b \bar b$ and $t\bar t t \bar t$ production
($t\bar t t\bar t$ is much smaller than $t \bar t b \bar b$ and so we 
ignore it in the rest of the paper.)
We study the feasibility of searching for the Higgs boson using
$Wh, Zh \to \ell^\pm \;(\ell = e, \mu)\; + 4b + X$ at the LHC.  A naive
signal analysis at the Tevatron already tells
us that the signal rate is too small for realistic detection.
At the LHC, we found a sufficiently large signal rate with a relatively
small background for $m_h \alt 160 $ GeV.
Reconstructing the invariant
mass of the 4 $b$-tagged jets is shown to play a crucial role:
The signal will peak at $m_h$ while the serious
background begins at $M_{4b} \agt 160$ GeV.  

Details of the Higgs sector of NMSSM
and SLH$\mu$ model are referred to
Refs.\,\cite{higgs} and \cite{simple}, respectively.
The dominant production for an intermediate Higgs boson at the LHC is the
gluon fusion. However, as mentioned above 
the decay $h \to \eta\eta$ followed by
$\eta \to b\bar b$ is overwhelmed by QCD backgrounds\,\cite{scott}.
The next production mechanism, the $WW$ fusion,
has the final state consisting of only hadronic jets.
Therefore, we consider
the associated production with a $W$ or $Z$ boson.
The cross section is proportional to the square of the coupling $g_{VVh}$.
In the NMSSM, the deviation of $g_{VVh}$ from the SM value 
depends on the nature of the $h_1$.
For the bench-mark points \#2 and \#3 of Ref.\,\cite{ellw}
the size of $g_{VVh}$ is very close to the SM value, though the sign
may be opposite.
We consider 2 bench-mark points A and B, which are very similar to
the bench-mark points \#2 and \#3 of Ref.\,\cite{ellw}, by scanning the
parameter space using NMHDECAY \cite{nmhdecay}.
In the SLH$\mu$, $g_{VVh}$ deviates from the SM value as
\begin{eqnarray}
 \frac{g_{WWh}^{\rm SLH}}{ g_{WWh}^{\rm SM} } &=& 1 - \frac{v^2}{3 f^2}
\left( \frac{s_\beta^4}{c_\beta^2} + \frac{c_\beta^4}{s_\beta^2} \right )
 + O \left ( \frac{v^4}{f^4} \right ) \nonumber 
 \\
 \frac{g_{ZZh}^{\rm SLH}}{ g_{ZZh}^{\rm SM} } &=& 1 - \frac{v^2}{3 f^2}
\left( \frac{s_\beta^4}{c_\beta^2} + \frac{c_\beta^4}{s_\beta^2} \right )
 - \frac{v^2}{4 f^2} ( 1 - t_W^2)^2 \nonumber \\
&& + O \left ( \frac{v^4}{f^4} \right ) \;, 
\end{eqnarray}
where $t_W$ is tangent of the Weinberg angle, $f$ is the symmetry breaking 
scale at TeV, $ c_\beta = \cos\beta , \, s_\beta=\sin\beta$, and
$\tan\beta$ is the ratio of the VEV of the two pseudo-Nambu-Goldstone
multiplets of the SLH$\mu$ model\,\cite{simple,song}.

We employ full helicity decays of the gauge bosons, 
$W \to \ell \nu$ or $Z\to \ell\ell$, and the phase decays of the Higgs boson
and the pseudoscalar in $h \to \eta \eta \to b \bar b b \bar b$.  
The detection requirements on the charged lepton and $b$ jets
in the final state are
\begin{eqnarray}
\label{cuts}
 p_T(\ell ) > 15 \, {\rm GeV}, && |\eta(\ell) | < 2.5\,,  \\
 p_T(b) > 15 \, {\rm GeV}, && |\eta(b) | < 2.5\,, \;\; \nonumber
   \Delta R (bb,b\ell) > 0.4\,,
\end{eqnarray}
where $p_T$ denotes the transverse momentum, $\eta$ denotes the pseudorapidity,
and $\Delta R = \sqrt{ (\Delta \eta )^2 + (\Delta \phi)^2}$ denotes
the angular separation of the $b$-jets and the lepton.
The smearing for the $b$ jets is
$
\frac{\Delta E}{E} = \frac{0.5}{\sqrt{E}} \oplus 0.03\,,
$
where $E$ is in GeV.
In order to minimize
the reducible backgrounds, we require to see
at least one charged lepton and 4 $b$-tagged jets in the final state.
We employ a $B$-tagging efficiency
of $70\%$ for each $B$ tag, and a probability of $5\%$ for a light-quark
jet faking a $B$ tag.

It is
possible for the photon in $\gamma +n j$ background to fake an electron
in the EM calorimeter.  However, we will ignore this
since the charged lepton from
the $W$ or $Z$ decay is quite energetic and produces
a track in the central tracking device, in contrast to that from a photon.
The backgrounds from $W + nj$ and $Z+nj$ 
contribute at a very low level and are reducible
as we require 4 $b$-tagged jets in the final state.
The background from $WZ \to \ell \nu b\bar b$
is also reducible by the 4 $b$-tagging requirement.
So is
QCD production of $t\bar t$ pair with one of the top decay hadronically
and the other semi-leptonically.  Jets from the $W$ decay may fake
a $B$-tag.  This background is under control after
applying our selective cuts.
While most of the backgrounds are reducible, there are a few channels that
are irreducible.
They are (i) $t \bar t b \bar b$ production, and (ii) $W/Z + 4 b$ production.

\section{Results}
\label{sec2}
As mentioned in the Introduction, we use two popular models for new physics:
(i) NMSSM and (ii) SLH$\mu$.
In NMSSM, we scan the code NMHDECAY\,\cite{nmhdecay}
and choose two bench-mark points, A and B, both of which have
$B(h \to a_1 a_1)\approx 1$ and $B(a_1 \to b\bar b) \approx 0.9$.
In a large portion of the
parameter space of NMSSM, the mass of $h_1$ is around 100 GeV and
$B(h_1 \to a_1 a_1) \agt 0.7$\,\cite{nmssm}.
The bench-mark points that we employ
are quite common in the NMSSM.
In the SLH$\mu$ model, we employ two
points in the parameter space such that the mass of the Higgs boson
is $\mathcal{O}(100)$ GeV and $B(h \to \eta \eta) \agt 0.7$\,\cite{song}.

We show the signal cross sections of $Wh$ and $Zh$
for the NMSSM and for SLH$\mu$ in Table
\ref{table1}, and various backgrounds in
Table \ref{table3}, respectively.
The cross sections are under the cuts listed in Eq.\,(\ref{cuts}).
We have imposed a $B$-tagging efficiency of $0.7$ for each $b$ jet and
a mis-tag efficiency of $0.05$ for a light-quark jet to fake a $b$ jet.
We require to see at least one charged lepton and 4 $b$-tagged jets.
We also show various couplings relative to the SM values in Table
\ref{table1}. With these values one can easily understand
the relative importance in various channels.  The quantity $C^2_{4b}$
defined by
\begin{equation}
 C^2_{4b} = \left( \frac{ g_{VVh}}{ g_{VVh}^{\rm SM} } \right )^2 \,
            B( h \to \eta\eta) \, B^2 ( \eta \to b\bar b)
\end{equation}
shows very clearly the importance of the channel $h \to \eta\eta\to
b\bar b b\bar b$ that we are considering.  For example, the two NMSSM
bench-mark points have $C^2_{4b} > 0.8$ while those for
SLH$\mu$ only have $C^2_{4b} \simeq 0.1$.  This explains why the
significance of the SLH$\mu$ signals is much smaller than that of
the NMSSM signals, shown in Table \ref{table4}.
The LEP Coll.\,\cite{delphi} has made model-independent searches for the
Higgs bosons in extended models.  They put limits on the quantity
$C^2_{4b}$ using the channel $e^+ e^- \to Zh \to Z A A \to Z + 4b$.  The
bench-mark points listed in the Tables are consistent with
the existing limits.

\begin{table}[b!]
\centering
\caption{\small \label{table3}
Various background cross sections under the same cuts and efficiencies
as in Table \ref{table1}.
}
\smallskip
\begin{tabular}{ll}
\hline
  Channels & cross sections (fb) \\
\hline
 $t\bar t$  &  172  (NMSSM \& SLH$\mu$) \\
 $t\bar t b\bar b$ & 236 (NMSSM), 284 (SLH$\mu$ A),
   429 (B) \\
 $W + 4b$ & 3.80 (NMSSM), 4.16 (SLH$\mu$ A), 4.63 (B)\\
 $Z + 4b$ & 3.85 (NMSSM \& SLH$\mu$) \\
\hline
\end{tabular}
\end{table}

\begin{figure}
\includegraphics[width=3.6in]{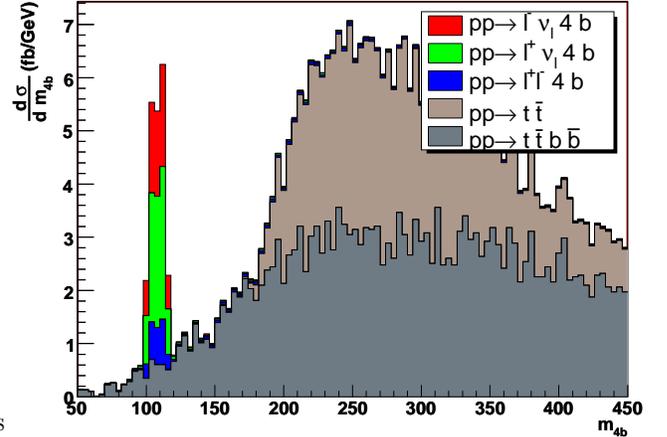}
\caption{
Invariant mass spectrum $M_{4b}$ of the signal and various backgrounds
for the bench-mark point B of the NMSSM. 
}
\label{fig1}       
\end{figure}

A comment on the background rates in Table \ref{table3} is in order here.
In general, one defines the background as in the SM.  However, here
we define the background for our search in $Wh, Zh \to \ell + 4b$
as those arising from the new physics under consideration.
The background in the NMSSM (including NMSSM interactions)
is the same as in the SM.
In the SLH$\mu$ model, however,
especially the $t\bar t b\bar b$ from $t\bar t \eta \to
t\bar t b\bar b$ increases the background substantially.
Suppose that the SLH$\mu$ is the actual model describing our world.
If we are searching for
the Higgs decay into pseudoscalar bosons, we have to fight against
the $t\bar{t}\eta \to t\bar t b\bar b$ background in the SLH$\mu$
model itself.
Nevertheless, if we look at combination of signal channels, this
$t\bar t b\bar b$ would be an interesting one for the $\eta$
boson.

Since we require all 4 $b$-tagged jets, we can reconstruct the
invariant mass $M_{4b}$ of the signal and the background.  We show the
invariant mass spectrum for the NMSSM point B in Fig. \ref{fig1}.  The
spectrum for other bench-mark points are similar.
For $m_h \alt
160$ GeV the signal peak will stand out of the continuum,
provided that the $B(h\to \eta \eta)$ still dominates.  We can calculate
the significance of the signal by evaluating the signal and background
cross sections under the signal peak:
\begin{equation}
m_h - 15 \;{\rm GeV} < M_{4b} < m_h + 15 \;{\rm GeV} \;,
\end{equation}
which is a conservative choice for the signal peak resolution.
We show the total signal and background cross sections and
the significance $S/\sqrt{B}$ in Table \ref{table4}
using an integrated luminosity of 30 fb$^{-1}$.  The significance of
the NMSSM bench-mark points are large because of the smallness of background.
On the contrary,
the SLH$\mu$ bench-mark points have smaller
significance
but close to 4 for point A, but not for point B.
  It is due to smaller signal rates and
a much larger background from $t\bar t \eta$ production.

\begin{table}[tb!]
\centering
\caption{\small \label{table4}
Total signal and background cross sections 
after applying the cuts in Eq.\,(\ref{cuts}) and
the invariant mass cut of
$ m_h - 15 \;{\rm GeV} < M_{4b} < m_h + 15 \;{\rm GeV}$.
The significance $S/\sqrt{B}$ is for a luminosity of 30 fb$^{-1}$.}
\smallskip
\begin{tabular}{lllll}
\hline
&  \multicolumn{2}{l}{NMSSM} & \multicolumn{2}{l}{SLH$\mu$} \\
       & A & B  &  A & B \\
\hline
signal & 6.53 fb & 18.85 fb  & 2.50 fb  & 1.25 fb  \\
\hline
bkgd & 4.83 fb  & 4.77 fb  & 13.83 fb  & 22.45 fb  \\
\hline
$S/\sqrt{B}$ & 16.3 & 47.3 & 3.7 & 1.4 \\
\hline
\end{tabular}
\end{table}

For $m_h$ below 250 GeV, the $t\bar t h$ cross section is subdominant
relative to the $Zh$ and $Wh$ production when the Higgs is SM-like.
In other models, however, the top Yukawa coupling can be much enhanced.
In this case, the $t\bar t h$ production could be dominant.
Unfortunately the signal analysis in $t\bar t h$ is more complicated because
of a total of 6 $b$ jets in the final state, but only 4 of those can be
reconstructed at $m_h$.  Therefore, efficiency will drop in picking the
right $b$ jets.

In conclusion, the dominance of $h\to \eta\eta$ decay mode is highly
motivated because it can relieve the little hierarchy problem and the
tension with the precision data.  However, the dominance of $h\to \eta\eta$
in the intermediate mass region worsens significantly the discovery
channels of $gg\to h \to \gamma \gamma$ and $Wh \to \ell\nu b\bar b$.
In this Letter, we have shown for the first time that $h\to\eta\eta \to 4b$
is complementary to make up for 
the loss of efficiencies in $h \to \gamma\gamma$ and $h\to b \bar b$ 
modes.   It is made possible by
considering the $Wh$ and $Zh$ production with at least one charged lepton
and 4 $B$-tags in the final state and we can identify a clean Higgs signal
with a full reconstruction of the Higgs boson mass.  Our work therefore
urges the experimenters to fully establish the feasibility of this mode.

The work was supported in part by the NSC of Taiwan 
and by Korea Research Foundation Grant.


\begin{thebibliography}{999}

\bibitem{prec}
  ALEPH, DELPHI, L3, OPAL, and SLD Collaborations,
  Phys.\ Rept.\  {\bf 427}, (2006) 257.

\bibitem{LEP}
  R.~Barate {\it et al.}  [LEP Working Group for Higgs boson searches],
  Phys.\ Lett.\ B {\bf 565}, (2003) 61.







\bibitem{nmssm}
R.~Dermisek and J.~F.~Gunion, Phys. Rev. Lett. {\bf 95}, (2005) 041801;
R.~Dermisek, J.~F.~Gunion and B.~McElrath, Phys. Rev. {\bf D76}, (2007) 051105;
R.~Dermisek and J.~F.~Gunion, arXiv:hep-ph/0611142.

\bibitem{song}
K.~Cheung and J.~Song, Phys. Rev. {\bf D76}, (2007) 035007.

\bibitem{simple}
M.~Schmaltz, JHEP {\bf 0408}, (2004) 056.

\bibitem{ours}
K. Cheung, J. Song and Q.S. Yan, Phys. Rev. Lett. {\bf 99}, (2007) 031801.

\bibitem{scott}
T.~Stelzer, S.~Wiesenfeldt and S.~Willenbrock, Phys. Rev. {\bf D75}, (2007)
077701.

\bibitem{carena}
U.~Aglietti {\it et al.}, arXiv:hep-ph/0612172.

\bibitem{4tau}
P.~W.~Graham, A.~Pierce and J.~G.~Wacker,  arXiv:hep-ph/0605162.

\bibitem{matchev}
B.~A.~Dobrescu, G.~Landsberg and K.~T.~Matchev, Phys. Rev. {\bf D63}, (2007)
075003 (2001);
S.~Chang, P.~J.~Fox and N.~Weiner, 
Phys.\ Rev.\ Lett.\  {\bf 98}, (2007) 111802.

\bibitem{assoc}
A.~Arhrib, K.~Cheung, T.~J.~Hou and K.~W.~Song, JHEP {\bf 0703}, (2007) 073.

\bibitem{higgs}
D.~J.~Miller, R.~Nevzorov and P.~M.~Zerwas,
 Nucl.\ Phys.\ B {\bf 681}, (2004) 3.

\bibitem{ellw}
U.~Ellwanger, J.~F.~Gunion and C.~Hugonie, JHEP {\bf 0507}, (2005) 041.

\bibitem{nmhdecay}
U.~Ellwanger and C.~Hugonie, Comput.\ Phys.\ Commun.\  {\bf 175}, (2006) 290.
U.~Ellwanger, J.~F.~Gunion and C.~Hugonie, JHEP {\bf 0502}, (2005) 066.

\bibitem{delphi}
S.~Schael {\it et al.}  [ALEPH, DELPHI, L3, and OPAL Collaboration],
  Eur.\ Phys.\ J.\  C {\bf 47}, (2006) 547;
J.~Abdallah {\it et al.}  [DELPHI Collaboration],
  Eur.\ Phys.\ J.\  C {\bf 38}, (2004) 1.

\end{thebibliography}
\end{document}